\documentclass[pra,twocolumn,superscriptaddress]{revtex4}
\usepackage[ansinew]{inputenc}
\usepackage[dvips]{graphicx}
\usepackage{amsmath}
\usepackage{layout}
\usepackage{float}
\usepackage{subfigure}
\usepackage{amsfonts}
\usepackage{amssymb}

\begin{document}

\title{Experimenter's Freedom in Bell's Theorem and Quantum Cryptography}
\begin{abstract}
Bell's theorem states that no local realistic explanation of quantum
mechanical predictions is possible, in which the experimenter has a freedom to
choose between different measurement settings. Within a local realistic
picture the violation of Bell's inequalities can only be understood if this
freedom is denied. We determine the minimal degree to which the experimenter's
freedom has to be abandoned, if one wants to keep such a picture and be in
agreement with the experiment. Furthermore, the freedom in choosing
experimental arrangements may be considered as a resource, since its lacking
can be used by an eavesdropper to harm the security of quantum communication.
We analyze the security of quantum key distribution as a function of the
(partial) knowledge the eavesdropper has about the future choices of
measurement settings which are made by the authorized parties (e.g.\ on the
basis of some quasi-random generator). We show that the equivalence between
the violation of Bell's inequality and the efficient extraction of a secure
key --- which exists for the case of complete freedom (no setting knowledge)
--- is lost unless one adapts the bound of the inequality according to this
lack of freedom.
\end{abstract}
\date{\today}

%

\author{Johannes Kofler}%

\affiliation
{Institut f\"{u}r Experimentalphysik, Universit\"{a}%
t Wien, Boltzmanngasse 5, 1090 Wien, Austria}%

\author{Tomasz Paterek}%

\affiliation
{Instytut Fizyki Teoretycznej i Astrofizyki, Uniwersytet Gda\'nski, PL-80-952 Gda\'nsk, Poland}%

\author{{\v C}aslav Brukner}%

\affiliation
{Institut f\"{u}r Experimentalphysik, Universit\"{a}%
t Wien, Boltzmanngasse 5, 1090 Wien, Austria}%

\affiliation
{Institut f\"ur Quantenoptik und Quanteninformation, \"Osterreichische Akademie der Wissenschaften,\\ Boltzmanngasse 3, 1090 Wien, Austria}%

\maketitle

\section{Introduction}

Can the experimenter's free-will be experimentally tested? It seems
unreasonable to expect that this question can be answered unconditionally. The
philosophical debate on whether or not our choices are ultimately up to
ourselves or are just a predetermined illusion has lasted for centuries
without reaching a final conclusion. Here we pursue our profession as
physicists reminding ourselves of Einstein's words that ''It is the theory
which decides what we can observe''. Thus, roughly speaking, one could say
that it is the theory that decides whether or not the experimenter's free-will
can be tested. Here we consider the experimenter's freedom of choosing between
different possible measurement settings and argue that, \textit{within a local
realistic theory}, it can experimentally be tested. Following Gill \textit{et
al}.~\cite{gill} this freedom will be defined as the \textit{independence} of
the experimenter's choice of measurement settings from the local realistic
mechanism that determines the actual measurement results.

The theorem of John Bell~\cite{bell} states the impossibility of a local
realistic explanation of quantum mechanics in which the experimenter has a
freedom to choose between different experimental arrangements. This is
demonstrated by the experimental confirmation of the violation of Bell's
inequalities in agreement with quantum mechanical predictions. The
philosophical implications of Bell's theorem are startling: either one must
abandon the experimenter's freedom, or the view that external reality exists
prior to and independent of observations (realism), or dramatically revise our
concepts of space and time (locality). Needless to say, either of the choices
requires radical revision of the ruling philosophical view of most of the
scientists and is in sharp contrast to our every-day experience.

If one chooses to keep a local realistic picture and to deny the
experimenter's freedom, then one must accept a world in which the measurement
settings that are generated by the experimenter's choice (or by tossing a
coin, or --- to put it in a grotesque way --- by the parity of the number of
cars passing the laboratory within $n$ seconds, where $n$ is given by the
fourth decimal of the cube of the actual temperature in degrees Fahrenheit)
are determined in advance and correlated with the actual outcome of the
measurement. Bell himself comments such theories with the words~\cite{bell1}:
''A theory may appear in which such conspiracies inevitably occur, and these
conspiracies may then seem more digestible than the non-localities of other
theories. When that theory is announced I will not refuse to listen, either on
methodological or other grounds. But I will not myself try to make such a theory.''

In this paper we show that the experimentally observed degree of violation of
Bell's inequalities sets a minimal degree to which the free choice has to be
abandoned if one insists on a local realistic explanation. Thus, not every
local realistic theory denying free choice is in agreement with observations.
Finally, extending the idea of Hwang~\cite{hwang}, we argue that the
experimenter's freedom can be considered as a resource in quantum
communication as its lacking can be used by an eavesdropper to harm its
security. We analyze the security in quantum key distribution as a function of
the knowledge the eavesdropper has about the choice of measurement settings
which is made by the authorized parties in the protocol.

\section{Bell's Inequalities with Reduced Experimenter's Freedom\label{sect}}

\textit{Realism} supposes that the measurement results are determined by
''hidden variables'' which exist prior to and independent of observation.
\textit{Locality} supposes that the results obtained at one location are
independent of any measurements or actions performed at space-like separated
regions. Finally, \textit{''freedom of choice''} assumes that the
experimenter's choice of the measurement setting is independent of the local
realistic mechanism which determines the measurement results. In what follows
we pursue the approach of Gill \textit{et al.}~\cite{gill} in formulating
these concepts in a mathematically rigorous way.

Consider two spatially separated partners, Alice and Bob, performing
space-like separated experiments on particles which are pairwise emitted by
some source. Let $X$ and $Y$ denote the \textit{actual} measurement outcomes
obtained, and $k$ and $l$ the actual measurement settings chosen by Alice and
Bob, respectively. The outcomes $X$ and $Y$ can take values $+1$ or $-1$, and
the settings $k$ and $l$ values $1$ or $2$. The probability to observe the two
outcomes to be equal, $X\!=\!Y$, under the chosen $k$ (Alice) and $l$ (Bob) is
denoted by $P(X\!=\!Y|kl)$.

Local realism assumes the existence of a quadruple of variables $\{X_{1}%
,X_{2},Y_{1},Y_{2}\}$, each taking values $+1$ or $-1$, which represents the
\textit{potential} measurement outcomes in a thought experiment, under any of
the possible measurement settings. This quadruple exists independently of
whether any or which experiment is actually performed on either side. Because
of locality the variables on Alice's side do not depend on the choice of
setting on Bob's side, and vice versa. Thus, local realism requires
$X\in\{X_{1},X_{2}\}$ and $Y\in\{Y_{1},Y_{2}\}.$

The freedom assumption expresses the independence between the choice $k,l$ of
the measurement settings and the local realistic mechanism which finally
selects the actual outcomes ${X,Y}$ from the potential ones ${X_{1}%
,X_{2},Y_{1},Y_{2}}$. Gill \textit{et al.}~\cite{gill} put this formally in
the requirement that $\{k,l\}$ are statistically independent of $\{X_{1}%
,X_{2},Y_{1},Y_{2}\}$. This means that in many thought repetitions of the
experiment the probabilities with which the quadruple $\{X_{1},X_{2}%
,Y_{1},Y_{2}\}$ takes on any of its $2^{4}$ possible values remain the same
within each subensemble defined by the four possible combinations of $k$ and
$l$. In particular, one has $P(X_{k}\!=\!Y_{l})\!=\!P(X\!=\!Y|kl)$, where
$P(X_{k}\!=\!Y_{l})$ is the (mathematical) probability for having $X_{k}%
=Y_{l}$.

What if the experimenter's freedom is just an illusion? Imagine that the
choices of experimental settings and experimental results are both
consequences of some common local realistic mechanism. In such a case the two
probabilities $P(X_{k}\!=\!Y_{l})$ and $P(X\!=\!Y|kl)$ may differ from each
other and we use their difference
\begin{equation}
\Delta_{kl}\equiv P(X\!=\!Y|kl)-P(X_{k}\!=\!Y_{l}) \label{measure}%
\end{equation}
to measure the lack of freedom. This measure can acquire values from $-1$ to
$1$, and the freedom case corresponds to all $\Delta_{kl}=0$. It is important
to note that while the probabilities $P(X\!=\!Y|kl)$ can directly be measured,
the $P(X_{k}\!=\!Y_{l})$ are only mathematical entities of the local realistic
theory without a direct operational meaning. Nevertheless, they satisfy a
set-theoretical constraint which is mathematically equivalent to the
Clauser--Horne--Shimony--Holt (CHSH) inequality~\cite{chsh}. The product of
local realistic results $X_{2}Y_{2}$ is always equal to the multiplication of
$(X_{1}Y_{1})(X_{1}Y_{2})(X_{2}Y_{1})$, because the square of a dichotomic
variable, with values $+1$ or $-1$, is equal to $+1$. This implies that the
following expression can attain only one of two values \cite{gill}:%
\begin{align}
&  \openone\{X_{1}=Y_{1}\}+\openone\{X_{1}=Y_{2}\}\nonumber\\
&  +\openone\{X_{2}=Y_{1}\}-\openone\{X_{2}=Y_{2}\}=0\text{ or }2\,,
\end{align}
where $\openone\{X_{k}=Y_{l}\}$ is the indicator of the event $X_{k}=Y_{l}$,
i.e., it is equal to $1$ if it happens and $0$ if it does not happen. The
expectation value of the indicator variable is the probability for the event
to happen, $P(X_{k}\!=\!Y_{l})$. Finally the expectation value of the
left-hand side cannot be greater than the maximum value of the averaged
expression:%
\begin{align}
S_{\text{CHSH}}  &  \equiv P(X_{1}\!=\!Y_{1})+P(X_{1}\!=\!Y_{2})\nonumber\\
&  \quad+P(X_{2}\!=\!Y_{1})-P(X_{2}\!=\!Y_{2})\leq2\,. \label{kids}%
\end{align}
The equivalence to the CHSH inequality is evident as soon as one recalls that
the correlation function of dichotomic variables equals $E_{kl}=2P(X_{k}%
\!=\!Y_{l})-1$. The above inequality, in turn, implies a new bound on the set
of probabilities that can experimentally be measured:%
\begin{align}
S_{\Delta}  &  \equiv P(X\!=\!Y|11)+P(X\!=\!Y|12)\label{delta}\\
&  \quad+P(X\!=\!Y|21)-P(X\!=\!Y|22)\leq2+\Delta_{\text{CHSH}},\nonumber
\end{align}
where $\Delta_{\text{CHSH}}\equiv\Delta_{11}+\Delta_{12}+\Delta_{21}%
-\Delta_{22}$. Note that on the basis of measured probabilities (relative
frequencies) one cannot make statements about the individual measures
$\Delta_{kl}$ but rather on their combination as given in $\Delta
_{\text{CHSH}}$. In particular, it is possible that $\Delta_{\text{CHSH}%
}\!=\!0$, although all the individual $\Delta_{kl}\!\neq\!0$, and it may also
be negative. However, only the case of positive $\Delta_{\text{CHSH}}$ ---
implying at least one individual $\Delta_{kl}$ to be unequal to zero --- makes
the freedom assumption within a local realistic model experimentally testable,
as the bound on the right-hand side of (\ref{delta}) is increased. As well one
could study the lower bounds of $S_{\text{CHSH}}$ and $S_{\Delta}$.

To give an example of the lack of freedom model, imagine a local realistic
mechanism in which the source ''knows'' in advance the settings ``to be
chosen'' by Alice and Bob. The source can ''arbitrarily'' manipulate the value
of $S_{\Delta}$ in this case. Even the algebraic (logical) bound of
$S_{\Delta}=3$ can be reached: whenever Alice and Bob both measure the second
setting, the source sends (local realistic) correlated pairs such that the
measurement results anticoincide, i.e.\ $P(X\!=\!Y|22)=0$, and in all other
measurements it produces pairs for which the results coincide,
i.e.\ $P(X\!=\!Y|11)=P(X\!=\!Y|12)=P(X\!=\!Y|21)=1$, and thus $S_{\Delta}=3$.
For this local realistic model (without freedom) inequality (\ref{delta}) is
satisfied, but only because of the adapted bound $2+\Delta_{\text{CHSH}}=3$.
Now imagine another experiment, in which the observers (freely) choose their
settings independently from the local realistic source. Then
$P(X\!=\!Y|kl)=P(X_{k}\!=\!Y_{l})$, i.e.\ $\Delta_{\text{CHSH}}=0$, and
inequality (\ref{delta}) is fulfilled with the bound of 2, as it becomes the
CHSH inequality (\ref{kids}).

The value of $\Delta_{\text{CHSH}}$ for which the inequality is still
satisfied, defines the minimal extent to which the experimenter's freedom has
to be abandoned such that a local realistic explanation of the experiment is
still possible. Denote the left-hand side of inequality~(\ref{delta}) as the
CHSH expression. The maximal possible quantum value, $S_{\text{QM}}=1+\sqrt
{2}$, of this expression can be observed for the maximally entangled state,
for example, the singlet state $|\psi^{-}\rangle=(|0\rangle|1\rangle
-|1\rangle|0\rangle)/\sqrt{2}$, where $|0\rangle$ and $|1\rangle$ are two
orthogonal quantum states, and for an appropriate choice of possible settings
$\{k,l\}$. This quantum value requires an abandonment of the experimentalist's
freedom to the extent of at least $\Delta_{\text{CHSH}}=\sqrt{2}%
-1\approx0.414$.

Since, based on the experiment, we can only make statements about
$\Delta_{\text{CHSH}}$, a large number of local realistic theories are
possible that deny the experimenter's freedom and are in agreement with
quantum mechanical predictions and experiments. In order to be able to make
further statements about these theories we need to impose some structure on
them. In what follows we restrict ourselves to the case where the degree to
which the freedom is abandoned --- that is the absolute value of the measure
$\Delta_{kl}$ --- is independent of the actual experiment performed,
i.e.\ $|\Delta_{kl}|\!=\!\Delta$ is the same for all $k,l$. Roughly speaking,
the level of conspiracy is assumed to be the same for all experimental
situations. Choosing $\Delta_{11}\!=\!\Delta_{12}\!=\!\Delta_{21}%
\!=\!-\Delta_{22}\!\equiv\!\Delta$ one obtains $\Delta=\frac{1}{4}(\sqrt
{2}-1)\approx0.104$ for the minimal degree required to explain the quantum
value of the CHSH expression by a local realistic model. If all the
$\Delta_{kl}$ are positive (i.e.\ if $P(X\!=\!Y|kl)>P(X_{k}\!=\!Y_{l})$ for
all $k,l$), one finds the even higher value $\Delta=\frac{1}{2}(\sqrt
{2}-1)\approx0.207$.

It is known that with an increasing number of parties, $N$, the discrepancy
between the results of Bell tests and local realistic predictions that respect
the experimenter's freedom increases rapidly (exponentially) with
$N$~\cite{mermin}. We now determine how the degree of the lack of freedom
needs to scale with $N$ in a local realistic theory that agrees with these tests.

Consider $N$ space-like separated parties who can each choose between two
possible measurement settings. Let $X^{(j)}\!\in\!\{1,-1\}$ denote the actual
measurement result obtained and $k_{j}\!\in\!\{1,2\}$ the actual measurement
setting chosen by party $j$. The probability to observe correlation, i.e.\ the
probability that the product of local results is equal to $1$ if settings
$k_{1},...,k_{N}$ are chosen, is denoted by $P(\prod_{j=1}^{N}\!X^{(j)}%
\!=\!1|k_{1}...k_{N})$. Local realism assumes the existence of $2\,N$ numbers
$\{X_{1}^{(1)},X_{2}^{(1)},...,X_{1}^{(N)},X_{2}^{(N)}\}$, each taking values
$+1$ or $-1$ and representing the potential measurement outcomes of $N$
parties under any possible combination of their measurement settings. The
(mathematical) probability that the product of the potential outcomes is equal
to $1$ is denoted by $P(\prod_{j=1}^{N}\!X_{k_{j}}^{(j)}\!=\!1)$. Note again
that this probability cannot be measured experimentally.

We apply the approach used above to the present case of $N$ parties. We
introduce the difference%
\begin{equation}
\Delta_{k_{1}...k_{N}}\equiv P(\textstyle\prod_{j=1}^{N}\!X^{(j)}%
\!=\!1|k_{1}...k_{N})-P(\textstyle\prod_{j=1}^{N}\!X_{k_{j}}^{(j)}\!=\!1)
\label{cold}%
\end{equation}
to measure the lack of freedom of $N$ experimenters. The probabilities
$P(\prod_{j=1}^{N}\!X_{k_{j}}^{(j)}\!=\!1)$ satisfy a set-theoretical
constraint that is mathematically equivalent to the Mermin
inequality~\cite{mermin} (the particular form used here is
from~\cite{zukowski}):%
\begin{equation}
M\equiv\!\sum_{k_{1},...,k_{N}=1}^{2}\!s(k_{1},...,k_{N})\,P(\textstyle
\prod_{j=1}^{N}\!X_{k_{j}}^{(j)}\!=\!1)\leq B(N)\,, \label{cave}%
\end{equation}
where $s(k_{1},...,k_{N})\!=\!\sin\!\left[  (k_{1}+...+k_{N})\frac{\pi}%
{2}\right]  $ are coefficients taking values $0$, $+1$ or $-1$. The inequality
is bounded by $B(N)\!=\!\frac{1}{2}\!\left[  2^{\lfloor N/2\rfloor
}\!+\!2^{N/2}\sin(\frac{N\pi}{4})\right]  $, where $\lfloor x\rfloor$ is the
greatest integer less or equal to $x$. Using inequality~(\ref{cave}) and
definition~(\ref{cold}), one obtains a new inequality:%
\begin{align}
M_{\Delta}  &  \equiv\!\sum_{k_{1},...,k_{N}=1}^{2}\!s(k_{1},...,k_{N}%
)\,P(\textstyle\prod_{j=1}^{N}\!X^{(j)}\!=\!1|k_{1}...k_{N})\nonumber\\
&  \leq B(N)+\Delta_{\text{Merm}}, \label{you}%
\end{align}
where $\Delta_{\text{Merm}}\!=\!\sum_{k_{1},...,k_{N}=1}^{2}s(k_{1}%
,...,k_{N})\,\Delta_{k_{1}...k_{N}}$. Importantly, the probabilities entering
this inequality are measurable.

In a Bell experiment involving the maximally entangled $N$-party (GHZ) state
one observes $M_{\text{QM}}=\frac{1}{2}\left[  2^{N-1}+2^{N/2}\sin(\frac{N\pi
}{4})\right]  $ for the maximal possible value of the left-hand side of
inequality~(\ref{you}). This implies $2^{N-2}-2^{\lfloor(N-2)/2\rfloor}$ for
the minimal value of $\Delta_{\text{Merm}}=M_{\text{QM}}-B(N)$ that still
allows a local realistic explanation of the experiment. Suppose again that the
degree of the lack of freedom is independent of the measurement setting. With
an adequate choice of signs one has $\Delta_{k_{1}...k_{N}}\!=\!\Delta_{N}$
for $k$'s for which $s(k_{1},...,k_{N})\!=\!1$ and $\Delta_{k_{1}...k_{N}%
}\!=\!-\Delta_{N}$ for $k$'s for which $s(k_{1},...,k_{N})\!=\!-1$. This
results in $\Delta_{\text{Merm}}\!=\!2^{N-1}\Delta_{N}$. Finally, one obtains
that the degree to which the experimenter's freedom has to be abandoned in
order to have an agreement between local realism and Bell's experiments with
$N$ parties \textit{saturates exponentially fast} with $N$ as $\Delta
_{N}=\frac{1}{2}-\frac{1}{2^{\lfloor(N+1)/2\rfloor}}$. In the limit of
infinitely many partners $\Delta_{N}$ reaches the value of $\frac{1}{2}$. It
is remarkable that if the sign of all $\Delta_{k_{1}...k_{N}}$ is chosen
positive, there will be no way to obtain agreement between local realism and
the experimental results, since $\Delta_{N}$ would have to leave the range
from $-1$ to $+1$ in the limit of large $N$. The other argument which
invalidates all $\Delta_{k_{1}...k_{N}}$ to be positive involves only four
parties. In this case in the expression defined in (\ref{you}) the number of
probabilities with a positive sign is equal to the number of probabilities
with a negative sign. Thus, if all $\Delta_{k_{1}...k_{N}}$ are positive and
have the same value they cancel each other, i.e.\ $\Delta_{\text{Merm}}=0$,
and no explanation of the violation of the bound $B(N\!=\!4)$ is possible.

In this section we showed that quantum correlations for $N$ partners can be
explained within local realism only if both the number of measurement settings
in which the experimenter's freedom is abandoned (all $2^{N-1}$ combinations
of local settings entering the Mermin inequality) increases exponentially and
the degree of this abandonment saturates exponentially fast with $N$.
Furthermore, from the viewpoint of local realism that denies freedom, there is
no obvious reason why the quantum bound of the CHSH inequality is $1+\sqrt{2}$
and not, for example, the maximal possible logical bound of 3. In our opinion
these objections clearly show that the local realistic program goes squarely
against every effort for simple and sensible explanations of our observations.
In particular, in order to explain the violation of Bell's inequality within
local realism, one has to introduce purely theoretical and experimentally not
accessible entities such as the (mathematical) probabilities $P(X_{k}%
\!=\!Y_{l})$. This is against the spirit of Ockham's razor principle.

\section{Quantum Key Distribution with Reduced Experimenter's Freedom}

The violation of Bell's inequality by legitimate parties was found to be a
necessary and sufficient condition for their efficient extraction of a quantum
secret key~\cite{gisin,acin}, showing an appealing connection between secure
key distribution and the violation of local realism.

Apart from its fundamental meaning, the freedom to choose between different
measurement settings can be regarded as an important resource in quantum
secret key distribution. In particular, as recently shown by
Hwang~\cite{hwang}, an eavesdropper can both simulate the violation of Bell's
inequality and successfully eavesdrop, if the freedom in choosing the settings
by legitimate partners is abandoned. Effectively, one can assume that each
measurement device chooses its settings according to a pseudo-random sequence
that is installed in the device beforehand. Such a model of lack of freedom
allows the eavesdropper to know the algorithm generating the pseudo-random
numbers, at least to some extent, and correspondingly predict the future
measurement settings of the legitimate parties.

In what follows we will consider an Ekert-like protocol~\cite{ekert} --- a
combination of the BBM92 protocol~\cite{bennet} (which is an application of
the BB84 protocol~\cite{bb84} to entangled states) and a CHSH test --- and
analyze both the violation of Bell's inequality and the security of the key
distribution as a function of the amount of knowledge that the eavesdropper
Eve (E) has about the settings chosen by the legitimate parties Alice (A) and
Bob (B). We have chosen this combination of BBM92 and CHSH --- henceforth
denoted as the BBM--CHSH protocol --- because it allows to test the
relationship between Bell's theorem and secure quantum key distribution in a
single protocol. (In contrast to the original Ekert protocol with
non-orthogonal key establishing settings, there is a security proof for the
BBM92 attached to the error rate.)

\begin{figure}[t]
\begin{center}
\includegraphics{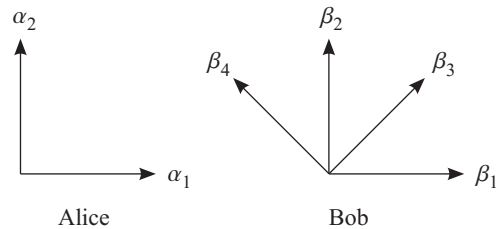}
\end{center}
\caption{Settings in the BBM--CHSH protocol. Alice chooses between two
orthogonal measurement directions $\alpha_{1}$ and $\alpha_{2}$, whereas Bob
has four different possibilities, namely the same directions as Alice, i.e.,
$\beta_{1}$ and $\beta_{2}$, as well as two directions rotated by $\frac{\pi
}{4}$, i.e. $\beta_{3}$ and $\beta_{4}$.}%
\label{fig_setting}%
\end{figure}

Consider a source that emits pairs of spin-$\frac{1}{2}$ particles in the
singlet state $|\psi^{-}\rangle=(|z+\rangle|z-\rangle-|z-\rangle
|z+\rangle)/\sqrt{2}$, where $|z+\rangle$ and $|z-\rangle$ denote spin-up and
spin-down along the $z$-direction, respectively. The legitimate parties
measure the incoming particles in the $x,y$-plane. Alice can choose between
two orthogonal settings, characterized by the azimuthal angles $\alpha
_{1}\equiv0$ and $\alpha_{2}\equiv\frac{\pi}{2}$, whereas Bob has 4 possible
measurement directions, namely $\beta_{1}\equiv\alpha_{1}\equiv0$, $\beta
_{2}\equiv\alpha_{2}\equiv\frac{\pi}{2}$, $\beta_{3}\equiv\frac{\pi}{4}$, and
$\beta_{4}\equiv\frac{3\pi}{4}$ (note that the $\beta_{j}$ are not numbered in
ascending order). Therefore, depending on their choice of settings, they
sometimes measure correlations for determining the violation of the CHSH
inequality, namely with the 4 settings $(\alpha_{1},\beta_{3})$, $(\alpha
_{1},\beta_{4})$, $(\alpha_{2},\beta_{3})$, and $(\alpha_{2},\beta_{4})$, or
they can establish a key, since their outcomes are perfectly anti-correlated
for measurements along $(\alpha_{1},\beta_{1})$ and $(\alpha_{2},\beta_{2})$.
If they choose $(\alpha_{1},\beta_{2})$ or $(\alpha_{2},\beta_{1})$,
i.e.\ orthogonal directions, they discard their results. A schematic of the
measurement directions is shown in figure~\ref{fig_setting}.

Let $P(X\!=\!-Y|ij)$ denote the probability that Alice and Bob obtain
anti-correlated results if they measure along $\alpha_{i}$ and $\beta_{j}$,
respectively, where $i=1,2$ and $j=1,2,3,4$. Within the freedom assumption the
(measured) CHSH expression has the form%
\begin{align}
S  &  \equiv P(X\!=\!-Y|13)+P(X\!=\!-Y|23)\nonumber\\
&  \quad+P(X\!=\!-Y|24)-P(X\!=\!-Y|14)\leq2\,. \label{chsh (bbm)}%
\end{align}
For the (maximally entangled) singlet state it is equal to $1+\sqrt{2}$. The
classical bound is 2, whereas the logical bound is equal to 3.

Let us now assume that an eavesdropper, Eve, has some knowledge about the
choice of settings of Alice and Bob, for instance by having some insight into
their random number generators. We \textit{model} this knowledge in the
following way: In each run, i.e., for each singlet pair, Eve knows that the
combination of local settings $(\alpha_{i},\beta_{j})$ will happen with
probability $q_{ij}$. For simplicity we assume that one out of the 8 joint
settings will happen with (high) probability $Q\geq\tfrac{1}{8}$, whereas all
the others 7 have equal (low) probability $\frac{1-Q}{7}$ to be manifested.
The number $Q$ shall be the same for all runs; the setting which it indicates
to be most probable of course changes from run to run. The case $Q=1$
corresponds to perfect knowledge of the eavesdropper and to the complete lack
of free will of Alice and Bob, whereas $Q=\tfrac{1}{8}$ means that Eve has no
knowledge at all.

Now we impose the following \textit{attack algorithm}: If Eve believes one of
the CHSH settings to be most likely, she sends the corresponding optimal
product state. In general, if $q_{ij}=Q$, which means that the setting
$(\alpha_{i},\beta_{j})$ is most probable from Eve's viewpoint, she intercepts
and sends either $\left|  \alpha_{i}\right\rangle _{\text{A}}\left|  \beta
_{j}+\pi\right\rangle _{\text{B}}$ or $\left|  \alpha_{i}+\pi\right\rangle
_{\text{A}}\left|  \beta_{j}\right\rangle _{\text{B}}$ (by tossing a fair
coin, such that the local results of Alice and Bob are always totally random).
Only in the special case $q_{14}=Q$, Eve sends $\left|  \alpha_{1}%
\right\rangle _{\text{A}}\left|  \beta_{4}\right\rangle _{\text{B}}$ or
$\left|  \alpha_{1}+\pi\right\rangle _{\text{A}}\left|  \beta_{4}%
+\pi\right\rangle _{\text{B}}$. This is the CHSH setting where the probability
of anti-correlation should be minimized, since $P(X\!=\!-Y|14)$ appears with a
minus sign in the CHSH inequality. Therefore, she attacks the CHSH
measurements in order to achieve a maximal violation ($S=3$) \textit{and} the
key establishing measurements to find the key (or rather produce it herself).

To further motivate why we have chosen this attack algorithm, we note that (i)
it is canonical in the way that Eve attacks all events in the same way, namely
with the appropriate product state. (ii) The attack is already good enough to
show that the connection between violation of local realism and secure key
distribution is lost in the case where the eavesdropper has partial knowledge
about the settings. (iii) Eve sends a product state for each pair that is
generated by the source; hence, Alice and Bob are faced with measurement
results that can be described by local realism but nevertheless can violate
the CHSH inequality (\ref{chsh (bbm)}) due to restricted freedom.

\begin{figure}[t]
\begin{center}
\includegraphics{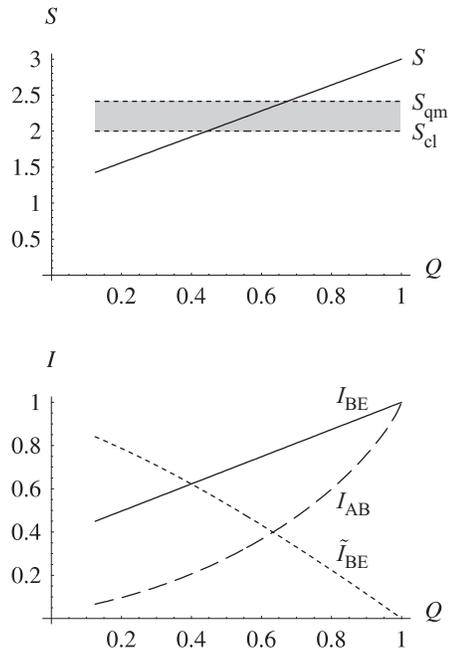}
\end{center}
\caption{(a) The (measured) CHSH expression $S$ as a function of Eve's setting
knowledge $Q$ (solid line). The CHSH inequality (\ref{chsh (bbm)}) with
classical bound $S_{\text{cl}}=2$ (dotted line) is violated for every setting
knowledge $Q>Q_{\text{cl}}\approx0.44$. The quantum bound $S_{\text{qm}%
}=1+\sqrt{2}$ is also indicated. (b) The mutual information between Alice and
Bob $I_{\text{AB}}$ (dashed line) and the actual mutual information between
Bob and Eve $I_{\text{BE}}$ (solid line), which is always smaller than (or
equal to) the Alice--Eve mutual information. For every setting knowledge $Q$
one has $I_{\text{BE}}\geq I_{\text{AB}}$ and thus Alice and Bob can never
extract a secret key. An optimal attack without setting knowledge leads to
$\tilde{I}_{\text{BE}}$ (dotted line). Only for $Q\leq Q_{0}\approx0.63$ the
BBM--CHSH protocol is secure, because Alice and Bob find $I_{\text{AB}}%
\leq\tilde{I}_{\text{BE}}$ and they will not use their key.}%
\label{fig_eve}%
\end{figure}

According to Eve's setting knowledge and the attack strategy, one can compute
the value for the CHSH expression as measured by Alice and Bob. \textit{In the
subensemble} of cases where, e.g., Alice measures along $\alpha_{1}$ and Bob
along $\beta_{3}$, Eve sends with probability $Q$ the product states resulting
in anti-correlations $P(X\!=\!-Y|13)=1$. In the rest of the cases she sends 7
possible ''wrong guesses'' which each happen with probability $\frac{1-Q}{7}$
and for each of them the probability for anti-correlations takes values
between $\frac{1}{2}$ and $\cos^{2}\!\frac{\pi}{8}\approx0.854$, depending on
the specific wrong attack. The measured probability $P(X\!=\!-Y|13)$ is the
expectation value of all 8 sets of anti-correlated results weighted with their
probabilities to happen. Analogously, the probabilities for anti-correlation
in the other subensembles are calculated and we find
\begin{align}
P(X\!  &  =\!-Y|13)=Q+\tfrac{1-Q}{7}\left(  \tfrac{5}{2}+2\,\cos^{2}%
\!\tfrac{\pi}{8}\right)  \!,\\
P(X\!  &  =\!-Y|23)=P(X\!=\!-Y|13)\,,\\
P(X\!  &  =\!-Y|24)=Q+\tfrac{1-Q}{7}\left(  \tfrac{5}{2}+\cos^{2}\!\tfrac{\pi
}{8}+\sin^{2}\!\tfrac{\pi}{8}\right)  \!,\\
P(X\!  &  =\!-Y|14)=\tfrac{1-Q}{7}\left(  \tfrac{5}{2}+\cos^{2}\!\tfrac{\pi
}{8}+\sin^{2}\!\tfrac{\pi}{8}\right)  \!.
\end{align}
The CHSH expression finally results in
\begin{equation}
S=3\,Q+\tfrac{1-Q}{7}\left(  5+4\,\cos^{2}\!\tfrac{\pi}{8}\right)
\approx1.2+1.8\,Q\,. \label{chsh}%
\end{equation}
Thus, the logical bound $S_{\text{log}}\equiv3$ is reached in the limit
$Q\rightarrow1$. The classical bound of $S_{\text{cl}}\equiv2$ is beaten for
all $Q>Q_{\text{cl}}\approx0.44$ and the quantum mechanics (Cirel'son) bound
$S_{\text{qm}}\equiv1+\sqrt{2}\approx2.41$ is beaten for setting knowledge
$Q>Q_{\text{qm}}\approx0.67$. If $Q$ is larger than $Q_{\text{qm}}$, Eve
should reduce the strength of her attack, e.g.\ by mixing some noise into her
product states, for otherwise even the quantum bound would be broken. The CHSH
expression (\ref{chsh}) and the bounds are shown in figure~\ref{fig_eve}%
a.

We can make the direct connection with section~\ref{sect}, where now Eve plays
the role of ''conspiratorial'' local realistic nature. Accordingly, we can
introduce the difference between the measured probabilities $P(X\!=\!-Y|ij)$
and their mathematical (set constraint fulfilling) counterparts $P(X_{i}%
\!=\!-Y_{j})$ to which the experimenter has no access. The latter are given by
the first if one substitutes the value $Q=\tfrac{1}{8}$ as this corresponds to
the case where Eve has no setting knowledge and Alice and Bob are receiving a
classical mixture of equally weighted product states for each run:
$P(X_{i}\!=\!-Y_{j})=P(X\!=\!-Y|ij)|_{Q=1/8}$. The difference%
\begin{equation}
\Delta_{ij}\equiv P(X\!=\!-Y|ij)-P(X_{i}\!=\!-Y_{j})
\end{equation}
measures the lack of freedom. The set constraint fulfilling CHSH expression
is
\begin{align}
S_{\text{CHSH}}  &  \equiv P(X_{1}\!=\!-Y_{3})+P(X_{2}\!=\!-Y_{3})\nonumber\\
&  \quad+P(X_{2}\!=\!-Y_{4})-P(X_{1}\!=\!-Y_{4})\leq2\,, \label{SCHSH}%
\end{align}
and therefore
\begin{equation}
S\leq2+\Delta_{\text{CHSH}}\,, \label{S(Q)}%
\end{equation}
where $\Delta_{\text{CHSH}}\equiv\Delta_{13}+\Delta_{23}+\Delta_{24}%
-\Delta_{14}$. Inequality (\ref{S(Q)}) is fulfilled if and only if
(\ref{SCHSH}) is fulfilled and this is the case because the (mathematical)
probabilities $P(X_{i}\!=\!-Y_{j})$ correspond to a mixture of product states
and therefore obey local realism.

When is Eve's knowledge about the settings also sufficient to find out the key
which is established by Alice and Bob? To answer this question, we have to
compute mutual informations between the parties. The mutual information
between Alice and Bob is determined by the bit error rate~\cite{gisin} which
they can compute in the \textit{subensembles} where they measured along
$\alpha_{1}\!=\!\beta_{1}\!=\!0$ or $\alpha_{2}\!=\!\beta_{2}\!=\!\frac{\pi
}{2}$. Let us consider the first; the error rate in the second is the same for
symmetry reasons. The bit error rate $D$ is given by the sum of 8 terms
corresponding to the 8 settings that were potentially possible from Eve's
point of view. Each term is the probability with which Eve believed this event
would happen --- $Q$ for the event $(\alpha_{1},\beta_{1})$ itself and
$\frac{1-Q}{7}$ for all the others (the wrong guesses), corresponding to our
definition of the setting knowledge --- multiplied with the probability that
the attack $(\alpha_{i},\beta_{j})$ leads to a correlation (error) rather than
an anti-correlation as for the original singlet state. This ''destruction
probability'' is $0$ for the ''correct'' event $(\alpha_{1},\beta_{1})$, it is
$\sin^{2}\!\frac{\pi}{8}$ for both $(\alpha_{1},\beta_{3})$ and $(\alpha
_{1},\beta_{4})$, and $\frac{1}{2}$ for all the others (where an orthogonal
state was sent to Alice or Bob). Finally, we find the bit error rate
\begin{equation}
D=\tfrac{1-Q}{7}\left(  \tfrac{5}{2}+2\,\sin^{2}\!\tfrac{\pi}{8}\right)
\approx0.4\,(1-Q)\,. \label{ber}%
\end{equation}
The mutual information between Alice and Bob is%
\begin{equation}
I_{\text{AB}}\equiv1-H(D)\,, \label{IAB}%
\end{equation}
with $H(p)\equiv-p\lg p-(1-p)\lg(1-p)$ the Shannon entropy, where lg denotes
the logarithm with base 2.

The maximal mutual information between Alice (or Bob for symmetry reasons) and
Eve from Alice's and Bob's viewpoint, which can be attained by an optimal
attack of Eve for a given error rate $D$ and \textit{under the condition that
Eve has no setting knowledge}, is given by~\cite{gisin}
\begin{equation}
\tilde{I}_{\text{AE}}=\tilde{I}_{\text{BE}}=1-H(\tfrac{1}{2}\!+\!\sqrt
{D-D^{2}})\,. \label{IAE-tilde}%
\end{equation}

The \textit{actual} mutual information between Alice and Eve, $I_{\text{AE}}$,
can be computed from the conditional entropy $H_{\text{A}|\text{E}}$ by
$I_{\text{AE}}=H_{\text{A}}-H_{\text{A}|\text{E}}$, where Alice's Shannon
information is $H_{\text{A}}=1$, since the outcomes of Alice are always
locally random for all possible attacks. As all chosen settings are publicly
revealed after the measurements, Eve can compute $H_{\text{A}|\text{E}}$ in
the \textit{subensemble} of the key establishing measurement $(\alpha
_{1},\beta_{1})$. (If Alice and Bob measure along $(\alpha_{2},\beta_{2})$,
the result does not change.) The calculation itself is straightforward, once
one realizes that $H_{\text{A}|\text{e}}=0$ for all 4 events e in which Eve
(justly) believed that Alice would choose $\alpha_{1}$, as Eve knows her
result in this case. If Eve made the (wrong) guess $\alpha_{2}$ then
$H_{\text{A}|\text{e}}=1$ for these 4 possible events, for Alice measures in
the orthogonal direction $\alpha_{1}$. Thus, $H_{\text{A}|\text{E}}%
=4\,\frac{1-Q}{7}$ and
\begin{equation}
I_{\text{AE}}=1-4\,\tfrac{1-Q}{7}=\tfrac{3}{7}+\tfrac{4}{7}\,Q\,. \label{IAE}%
\end{equation}
Analogously, one can find the \textit{actual} mutual information between Bob
and Eve:
\begin{equation}
I_{\text{BE}}=1-\tfrac{1-Q}{7}\left(  2+4\,H(\cos^{2}\!\tfrac{\pi}{8})\right)
\approx0.37+0.63\,Q\,,
\end{equation}
which is always smaller than (or equal to) $I_{\text{AE}}$. Secret-key
agreement between Alice and Bob using only error correction and privacy
amplification is possible if and only if the Alice-Bob mutual information is
greater than the minimum of the Alice--Eve and Bob--Eve mutual information,
that is, if and only if $I_{\text{AB}}>\min(I_{\text{AE}},I_{\text{BE}}%
)$~\cite{csiszar}. We have%
\begin{equation}
I_{\text{AB}}\leq I_{\text{BE}}%
\end{equation}
for all $Q$ and equality only holds for $Q=1$. Alice and Bob can never extract
a secret key, since the condition $I_{\text{AB}}>I_{\text{BE}}$ is never
fulfilled (figure~\ref{fig_eve}b).

The well-known critical error rate $D_{0}=\tfrac{1}{2}\,(1-\tfrac{1}{\sqrt{2}%
})\approx0.15$ corresponds, according to (\ref{ber}), to a setting knowledge
$Q_{0}\approx0.63$. For this knowledge $I_{\text{AB}}=\tilde{I}_{\text{BE}}$.
If $Q>Q_{0}$, the BBM--CHSH protocol is insecure, since $I_{\text{AB}}\leq
I_{\text{BE}}$ \textit{and} Alice and Bob find both their error rate to be
sufficiently small (below $D_{0}$) and the CHSH inequality (\ref{chsh (bbm)})
to be violated, which makes them think they are safe. For $Q\leq Q_{0}$ Eve's
setting knowledge is ''insufficient'' and the protocol becomes secure: Alice
and Bob cannot extract a secret key because still $I_{\text{AB}}<I_{\text{BE}%
}$, but they find $I_{\text{AB}}\leq\tilde{I}_{\text{BE}}$ and know that there
might be an eavesdropper and thus they will not use the key. For $0.44\approx
Q_{\text{cl}}<Q\leq Q_{0}\approx0.63$ they will find the CHSH inequality
(\ref{chsh (bbm)}) to be violated ($S>2$) and nonetheless they cannot extract
a secret key ($I_{\text{AB}}\leq\tilde{I}_{\text{BE}}$, $D\geq D_{0}$).
Therefore, we deduce that the equivalence between the violation of Bell's
inequality (with complete freedom) and the secure key distribution (without
freedom) is lost. (However, the new bound in inequality (\ref{S(Q)}) is never
broken and, in fact, a secret key can never be extracted.)

If Alice and Bob knew $Q$, which means they knew to which extent their freedom
is restricted, and if they calculated the maximal $\Delta_{\text{CHSH}}$ under
the constraint of an insecure key, $I_{\text{AB}}\leq\min(I_{\text{AE}%
},I_{\text{BE}})$, \textit{for all possible attacks}, then a violation of the
CHSH inequality with the new bound $2+\Delta_{\text{CHSH}}$ would be
equivalent to the possibility of efficient secret key extraction (unless the
new bound is larger than $S_{\text{qm}}=1+\sqrt{2}$). A violation of this new
bound is equivalent to statement that the classical bound 2 is violated in the
case of total freedom and for this situation there exists a complete
equivalence between the CHSH inequality violation and the security of the BBM
protocol~\cite{gisin,acin}.

\section{Conclusions}

The violation of Bell's inequalities is an experimental fact. Within a local
realistic program this fact can only be explained if the experimenter's
freedom in choosing between different measurement settings is denied (modulo
known loopholes, considered by most scientists to be of technical nature). For
a local realist our results show that both the number of settings in which the
freedom is abandoned grows exponentially and the degree of this abandonment
saturates exponentially fast with the number of parties. For the present
authors, however, these results are rather an indication of the absurdity of
the program itself. Nevertheless, they give rise to new security criteria for
quantum cryptography in situations in which the measurement settings chosen by
the authorized parties are partially revealed by an eavesdropper. This
contradicts the standard assumption in cryptography in which the laboratories
of the authorized parties are safe and no relevant information is allowed to
leak out from them. If this assumption is not fulfilled, we showed that the
violation of the standard CHSH inequality is not equivalent to a secure key
distribution anymore. Nevertheless, it is possible to define a new (higher)
bound whose violation indeed guarantees the security of the key. Therefore,
one can keep the security while, to some extent, relaxing the assumption that
no information is revealed to an eavesdropper, as long as the amount of this
information is known.

\section*{Acknowledgements}

This work has been supported by the Austrian Science Foundation (FWF) Project
SFB 1506 and the European Commission (RAMBOQ). {\v{C}}.\ B.\ thanks the
British Council in Austria. T.\ P.\ is supported by FNP and MNiI Grant No.\ 1
P03B 04927. The collaboration is part of an OeAD/MNiI program.

\end{document}